\documentclass[fleqn,10pt]{wlscirep}
\usepackage{hyperref}
\usepackage{float}
\usepackage[super]{nth}
\usepackage{multirow}
\usepackage{soul}
\title{Network Analysis of Uniswap: Centralization and Fragility in the Decentralized Exchange Market}

\author[1,2]{Tao Yan}
\author[1,2,*]{Claudio J. Tessone}

\affil[1]{University of Zurich, Blockchain \& Distributed Ledger Technologies, Department of Informatics, Z\"urich, Switzerland}
\affil[2]{UZH Blockchain Center, Z\"urich, Switzerland}
\affil[*]{claudio.tessone@uzh.ch}

\keywords{Blockchain, DeFi, Uniswap, Complex Networks, Centralization}

\begin{abstract}
The Uniswap is a Decentralized Exchange (DEX) protocol that facilitates automatic token exchange without the need for traditional order books. Every pair of tokens forms a liquidity pool on Uniswap, and each token can be paired with any other token to create liquidity pools. This characteristic motivates us to employ a complex network approach to analyze the features of the Uniswap market. This research presents a comprehensive analysis of the Uniswap network using complex network methods. The network on October 31, 2023, is built to observe its recent features, showcasing both scale-free and core-periphery properties. By employing node and edge-betweenness metrics, we detect the most important tokens and liquidity pools. Additionally, we construct daily networks spanning from the beginning of Uniswap V2 on May 5, 2020, until October 31, 2023, and our findings demonstrate that the network becomes increasingly fragile over time. Furthermore, we conduct a robustness analysis by simulating the deletion of nodes to estimate the impact of some extreme events such as the Terra collapse. The results indicate that the Uniswap network exhibits robustness, yet it is notably fragile when deleting tokens with high betweenness centrality. This finding highlights that, despite being a decentralized exchange, Uniswap exhibits significant centralization tendencies in terms of token network connectivity and the distribution of TVL across nodes (tokens) and edges (liquidity pools).
\end{abstract}
\begin{document}

\flushbottom
\maketitle
\thispagestyle{empty}

\section{Introduction}
Decentralized Finance (DeFi) has emerged as a popular domain in recent years, among the various components of DeFi, decentralized exchanges (DEXs) play a crucial role in facilitating token exchange and improving token liquidity. Uniswap stands out as one of the leading platforms in the DEX field, with approximately 4.1 billion USD locked on the platform as of December 10, 2023, according to DeFiLlama\cite{UniswapD66}. Unlike traditional exchanges that rely on order books for transaction matching, Uniswap employs a distinct mechanism called Automated Market Maker (AMM)\cite{xu2021sok}, where trading occurs directly with liquidity pools rather than with other individuals. These liquidity pools, also known as \textbf{pairs}, consist of a pair of tokens and are open to anyone to create using any two tokens. As a result, the Uniswap market forms a diverse network, wherein nodes represent tokens and edges represent liquidity pools. This excellent attribute enables the application of complex network methodologies for analysis. 

At the time of writing, Uniswap has 3 versions, the mainly used versions are version 2(V2)\cite{adams2020uniswap} and version 3(V3)\cite{adams2021uniswap}. As of October 31, 2023, there are 260,544 pairs created on Uniswap V2 and 15,350 pairs on Uniswap V3. Due to the larger token and liquidity pool network on Uniswap V2 compared to Uniswap V3, this study focuses on analyzing the Uniswap V2 market. Specifically, we examine both static and dynamic network properties from the perspective of a complex network. From a static perspective, we investigate the network's attributes on October 31, 2023, and find that this network exhibits scale-free and core-periphery properties, indicating that it is a fragile network. Additionally, our empirical analysis reveals that the degree distribution and TVL (Total Value Locked) distributions of tokens and liquidity pools in this network all conform to the power-law distribution, indicating the presence of token hubs and liquidity pool hubs in this network. Furthermore, we explore the dynamic features of the network and observe its increasing fragility. This fragility motivates us to consider the robustness of the network, especially in the face of extreme events like the Terra collapse\cite{badev2023interconnected}. Therefore, we conduct a robustness analysis by measuring the liquidity pools left, network connectivity and TVL loss upon four methods of removing nodes. The findings suggest that the Uniswap network showcases robustness in terms of removing tokens randomly, but it exhibits vulnerability particularly when removing tokens with significant betweenness centrality. 

The contributions of this paper can be summarized as follows:

\begin{itemize}
  \item First, this study represents the first comprehensive analysis of the entire Uniswap market from a complex network perspective. Our research uncovers that Although Uniswap is a decentralized exchange, it exhibits centralized characteristics in terms of network connectivity and TVL distribution.

  \item Second, we employ centrality metrics and robustness analysis from complex network theory to identify the tokens and liquidity pools that are crucial for maintaining network connectivity and liquidity.

  \item Finally, our research contributes to the broader field of complex networks by constructing the Uniswap market as a network in the DeFi field. More studies can be conducted on this network to identify arbitrage opportunities\cite{yan2024optimizing,zhang2024improved} on decentralized exchanges or to optimize the Uniswap router algorithm to swap tokens\cite{diamandis2023efficient,angeris2022optimal}.
\end{itemize}
The remainder of this paper is organized as follows. Section II provides crucial background information. Section III reviews related works that apply complex network methodologies to blockchain and cryptocurrency. Section IV details data collection methods. Section V outlines the methods employed in our analysis. Section VI presents the static features of the current Uniswap network, while Section VII introduces the dynamic features. Section VIII conducts the robustness analysis, and Section IX concludes the paper and discusses future research directions.

\section{Background}
\subsection{Automated Market Maker(AMM) Mechanism}
Automated Market Maker (AMM)\cite{xu2021sok} is a trading mechanism that facilitates automatic token exchange without the need for traditional order books. Uniswap is the first platform adopting the AMM by employing the Constant Product Formula, \(K = X \cdot Y\), to maintain liquidity pools, where \(X\) is the reserve of token A, and \(Y\) is the reserve of token B in the pool.
Users trade tokens with liquidity pools; anyone can create a liquidity pool, provide liquidity, or trade any amount of tokens on Uniswap. The smart contract automatically adjusts token prices based on the reserves of the two tokens in the liquidity pool. The price of tokens in the pool is determined by \(P_{A} = \frac{X}{Y}\) and \(P_{B} = \frac{Y}{X}\). 

\subsection{Liquidity and Total Value Locked (TVL)}
In traditional finance, liquidity\cite{liqudity_definition} in a share market usually refers to the ease with which a security can be bought or sold. The more liquid a security is, the more quickly and easily it can be bought or sold at a reasonable price. However, liquidity in the DeFi field is commonly assessed by the Total Value Locked (TVL), which represents the aggregate value of assets held or deposited within a DeFi protocol. Typically, TVL is denoted in USD. In the case of Uniswap, the TVL of a liquidity pool can be calculated using the formula:
\begin{equation}
\text{TVL}_{LP} = P_{A} \cdot R_{A} + P_{B} \cdot R_{B}\label{eq1}
\end{equation}
where \(P_{A}\) and \(P_{B}\) are the prices of token A and token B denoted in USD, and \(R_{A}\) and \(R_{B}\) are the reserves of token A and token B. The TVL of the entire Uniswap market can be calculated by
\begin{equation}
\text{TVL}_{Uniswap} = \sum_{i=1}^{n} \text{TVL}_{i}\label{eq2}
\end{equation}
where \(n\) is the number of liquidity pools on Uniswap and \(i\) represents a liquidity pool. The size and growth of TVL are often considered indicators of the popularity of a DeFi platform, reflecting user trust and engagement.
The TVL of a token can be calculated by 
\begin{equation}
\text{TVL}_{token} = \sum_{i=1}^{n} P \cdot R_{i}\label{eq3}
\end{equation}
where \(n\) is the number of liquidity pools paired with this token, \(P\) represents the price of this token in USD at a specific time and \(R_{i}\) denotes the token's reserve in liquidity pool \(i\). 

\section{Realted work}
The initial research that applies a complex network approach to blockchain and cryptocurrency transactions is conducted by \cite{2013An}. In their study, they examine the topological structure of both transaction and user networks and discover that these networks exhibit a non-trivial topology. Another study\cite{campajola2022evolution} involves constructing weekly networks for seven cryptocurrencies’ transactions, where the authors find that the degree distributions of these networks followed a power-law distribution. Besides the cryptocurrency transaction network, a comprehensive analysis of Ethereum leveraging graph analysis is conducted in \cite{chen2020understanding}, where smart contract creation and smart contract invocation are also used to build networks to analyze the characteristics of users and smart contracts. The Ether, Binance, USDT and Chainlink transaction networks are analyzed in 
\cite{de2021heterogeneous}, and it is shown that a few addresses rapidly become hubs in the transaction networks. A comprehensive analysis of Ethereum-based token transaction networks by \cite{de2022structural} demonstrates that smart contracts and exchange-related addresses play a fundamental role in the DeFi ecosystem, challenging the notion of complete decentralization in DeFi ecosystems.  This centralization trend is further supported by \cite{de2024patterns}, which reveals a rising centralization process in both ERC-20 and ERC-721 token networks. A daily ERC20 token trading graph is created in \cite{Pradeep_Dyapa_Jalan_Pradhan_2023} to observe the evolution of traders' trading activities and the dynamics of ERC20 tokens during the 2018 crypto crash and the COVID-19 pandemic. Additionally, \cite{li2023reward} investigates the fairness of Proof-of-Stake platforms and reveals a trade-off between system inclusiveness and reward distribution fairness, highlighting the complexity of achieving true decentralization in blockchain systems.

There are three papers that are closely related to our work. Firstly, the 50 largest liquidity pools on the Uniswap market are constructed as a network where tokens are nodes and edges are liquidity pools in \cite{lehar2021decentralized}. The 
core-periphery structure is found, and the majority of tokens are traded with WETH. Additionally, the Uniswap V2 and SushiSwap networks are created in \cite{mohammadi2023a}, where nodes represent tokens and edges signify the existence of at least one transfer between tokens. The two networks are compared based on different network feature metrics. Lastly, a transaction graph is constructed in \cite{Miori_Cucuringu_2022} to represent each liquidity taker on Uniswap V3. The nodes in this graph are the liquidity taker's executed transactions, and the edges represent the time elapsed between any two transactions. This network is utilized to examine the liquidity consumption behavior of market participants.

\section{Data description}
We collect the liquidity pool created, liquidity pool daily snapshot, and token price data from the \textbf{Uniswap subgraph}\cite{uniswap-v2-dev}. The liquidity pool created data is employed to establish the network, while the liquidity pool daily snapshot and token price data are mainly utilized to calculate the Total Value Locked (TVL) of both the liquidity pools and the tokens. The time range of this dataset spans from May 5, 2020 to October 31, 2023.
\subsubsection{Liquidity pool data}
We collect the liquidity pool data by retrieving all records in the query \textbf{pairs} within this subgraph.  One advantage of utilizing the Uniswap subgraph is the ability to combine detailed token information while collecting the liquidity pool data. The data collected encompasses the pair address, creation timestamp, name, symbol, and address of the two tokens in each pair. 
\subsubsection{Liqudity pool TVL data}
The total value locked (TVL) of liquidity pools in Uniswap's subgraph data structure is represented as '\text{reserveUSD}'. We collect the daily TVL of liquidity pools from 
the query \textbf{pairDayDatas}, where the data field contains \textit{pair id}, \textit{date}, \textit{reserve0}, \textit{reserve1}, and \textit{reserveUSD}. It is crucial to note that the pair TVL data obtained from the Uniswap subgraph may sometimes present excessively large values which are not correct, especially for the pairs in which both tokens are uncommon tokens. Therefore, it is necessary to identify and remove these outlier TVL values by comparing the tokens' prices on centralized exchanges. 

\subsubsection{Token TVL data}
Daily token price data is collected from the Uniswap subgraph query \textbf{tokenDayDatas}; where the data field includes the \textit{date}, \textit{token address} and \textit{daily price}. Originally, the TVL of a token can be calculated using equation \ref{eq3}, but because the price of some uncommon tokens obtained from the Uniswap subgraph is not accurate, it is not proper to use equation \ref{eq3} to calculate the TVL of these tokens. Although the price of some uncommon tokens may exhibit extreme values, the daily prices of widely used tokens such as WETH, USDC, USDT, DAI, and WBTC are highly reliable. By examining the number of liquidity pools that involve at least one of these five tokens, we observe that $98.34\%$ (262,402 / 266,826) of the liquidity pools are paired with one of these five tokens. Therefore, an alternative method for calculating the TVL of rare tokens can be employed as follows:



\begin{equation}
T V L_{\text {token }_{Aj}}=\sum_{j=1}^{n}\left(T V L_{L P_{j}}-P_{B}\cdot R_{B_{j}}\right)
\end{equation}

Here, $TVL_{\text {token }_{A}}$ denotes the TVL of ${token }_{A}$ throughout the market, and $n$ represents that ${token }_{A}$ has $n$ liquidity pools on the market, $T V L_{L P_{j}}$ represents the TVL of liquidity pool $j$, $B_{j}$ is another token in the liquidity pool $j$ besides token $A$, $P_{B}$ is the price in USD for token B. Token B represents one of the following tokens: WETH, USDC, USDT, DAI, and WBTC, while token A is another token in the pool. For liquidity pools not paired with these five tokens, the token TVL denoted as {\itshape reserveUSD} obtained from the Uniswap subgraph is used. After this processing, we rebuild the daily token value in a liquidity pool at a given day. The TVL of a token on a particular day is the sum of all the values of the token in all corresponding pairs.
\begin{equation}
TVL_{\text{token}_A} = \sum_{j=1}^{n} \left( TVL_{\text{token}_{Aj}} \right)
\end{equation}

\section{Methodology}\label{Method}
In this section, we present the methodology used in this study to analyze the characteristics of the Uniswap network from a complex network perspective. The overall goal is to investigate the structure and dynamics of the Uniswap market and identify important tokens and liquidity pools.
\paragraph{Network construction} 
The Uniswap network is represented as a weighted undirected graph $G = (V, E, W_{v}, W_{e})$, where:
\begin{itemize}
    \item $V$: Set of nodes that represent the addresses of the ERC-20 tokens.
    \item $E$: Set of edges indicating liquidity pools between two tokens.
    \item $W_{v}$: Set of properties of nodes, including the symbol, name, and token TVL.
    \item $W_{e}$: A set of edge properties, including names of liquidity pools, TVL, and creation timestamp.
\end{itemize}
This network shows excellent properties, where the nodes are tokens, and the edges are liquidity pools. We create the network on October 31, 2023 to examine the static characteristics of Uniswap V2. Additionally, we build all daily networks spanning from May 5, 2020, when Uniswap V2 was launched, up until October 31, 2023. Through this series of networks, we investigate the progressive daily dynamics. 

\paragraph{Power-law fitting}\label{Powerlaw_description}
The power-law is commonly used to describe scale-free distributions, it can be expressed as:
\begin{equation}
    P(x) = Cx^{-\alpha}
\end{equation}

where $P(x)$ is the probability of an event $x$, $C$ is a normalization constant, and $\alpha$ is the exponent of the power-law distribution. As ${\alpha}$ increases, the tail of the distribution decays more rapidly, leading to a concentration of the distribution around smaller values. Consequently, the probability of extreme events decreases. We employ the Python library \textbf{power-law}\cite{alstott2014powerlaw} to fit the degree distribution, token's TVL distribution and liquidity pool's distribution. The returns of the fitting function are $\alpha$, ${x_{min}}$, ${sigma}$, ${D}$, where $\alpha$ and ${x_{min}}$ represent the power-law exponent and the lower cut-off, respectively, ${sigma}$ is the standard deviation of the logarithmic binned data, and ${D}$ is the Kolmogorov-Smirnov distance\cite{massey1951kolmogorov}.


\paragraph{Average degree} 
The average degree is equal to the total degree divided by the number of nodes in the network. It can be used to determine the sparsity and tightness of the network.
For a node $i$, its degree $k_i$ is calculated as:
\begin{equation}
     k_i = \sum_{j} A_{ij}  
\end{equation}

where $A_{ij}$ is the adjacency matrix element, with a value of 1 or 0, where 1 indicates the presence of an edge between nodes $i$ and $j$.

For a graph $G$, the average degree $\langle k \rangle$ of the entire network is calculated as\cite{barabasi2013network}:
\begin{equation}
    \langle k \rangle = \frac{1}{N} \sum_{i} k_i
\end{equation}

where $N$ is the number of nodes in the graph, and $k_i$ is the degree of node $i$.

\paragraph{Node betweenness centrality} 
The betweenness of a node $v$ is the sum of the fraction of all-pairs shortest paths that pass through $v$.
Betweenness centrality of a vertex $C_{B}(v)$ is defined as\cite{brandes2001faster},
\begin{equation}
C_{B}(v)=\sum_{\substack{s \neq v \neq t \in V \\ s \neq t}} \frac{\sigma_{s t}(v)}{\sigma_{s t}}
\end{equation}
where ${\sigma_{s t}}$ is the number of shortest paths from $s$ to $t$, and ${\sigma_{s t}(v)}$ is the number of shortest paths from $s$ to $t$ that pass through a vertex $v$. The node betweenness centrality values are usually normalized by dividing through the number of pairs of vertices not including  $v$, which is  $(n-1)(n-2) / 2$ for undirected graphs.
\paragraph{Edge betweenness}
The betweenness of an edge $e$ is the sum of the fraction of all-pairs shortest paths that pass through $e$\cite{brandes2008variants},
\begin{equation}
C_{B}(e)=\sum_{s, t \in V} \frac{\sigma(s, t \mid e)}{\sigma(s, t)}
\end{equation}
where $V$ is the set of nodes, $\sigma(s, t)$ is the number of shortest paths, and $\sigma(s, t \mid e)$ is the number of those paths passing through edge $e$. The edge betweenness values are normalized by dividing $n(n-1)/ 2$ for undirected graphs.

\paragraph{k-core decomposition} 
The k-core of a network represents a highly connected region. We use the following formula to calculate the k-core of the network\cite{batagelj2003m}:
\begin{equation}
k_{core} = \arg\min_k \left( \frac{{\sum_{i=1}^n k_i}}{{n}} \geq k \right)
\end{equation}
where \(k_i\) represents the degree of node \(i\), \(n\) represents the total number of nodes in the network, and \(k\) is an integer, symbolizing the desired level of coreness.

\paragraph{Gini Coefficient}
The Gini coefficient\cite{yan2024analyzing} is used to measure the inequality of a distribution. We calculate the Gini coefficient of TVL for tokens and liquidity pools to understand its concentration in the network. The Gini coefficient is calculated by the following formula:

\begin{equation}
G=\frac{\sum_{i=1}^{n} \sum_{j=1}^{n}\left|x_{i}-x_{j}\right|}{2 n \sum_{i=1}^{n} x_{i}}
\end{equation}

where \(n\) represents the total number of tokens or liquidity pools, \(x_i\) and \(x_j\) represent the TVL of two respective tokens or liquidity pools.


\section{Static feature of the current Uniswap Network}
We conduct a static analysis of the Uniswap market on October 31, 2023. The market is represented as a graph, where tokens are nodes and liquidity pools are edges. Nodes and edges have their own attributes, such as address, total value locked (TVL), and creation timestamp. We investigate this network by analyzing the degree distribution and TVL distribution of both tokens and liquidity pools, and we employ k-core decomposition to identify the tightly connected groups within this market. Furthermore, we determine the most significant tokens and liquidity pools by utilizing the node betweenness centrality and edge betweenness centrality.
\subsection{Scale-free property}
We first analyze the degree distribution of tokens. Figure \ref{fig:Degree_Distribution} illustrates the Probability Density Function (PDF) and a power-law fitted function with an ${\alpha}$ value of 2.55. Furthermore, it shows a low Kolmogorov-Smirnov distance (D) of 0.03. The results reveal that the degree distribution of this Uniswap network follows the Power-law distribution when the degree {$k\geq 3$}, indicating that there are a few nodes with a large number of connections, while the majority of nodes have a small number of connections. Therefore, we can conclude that it is a scale-free network. 
\subsection{Liquidity centralization}
\begin{figure}[tbh]
  \centering
  \includegraphics[width=0.7\linewidth]{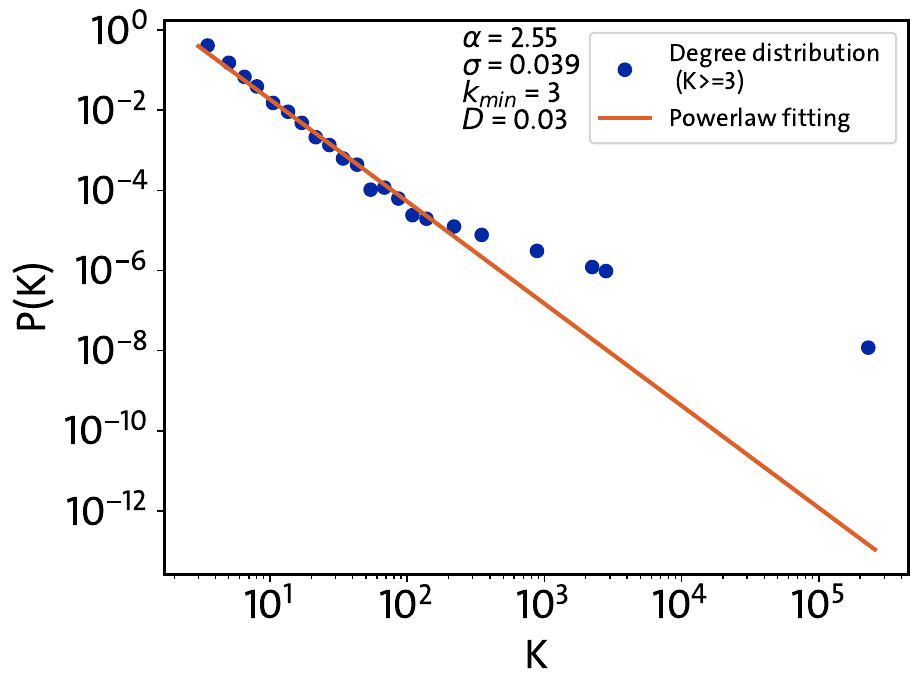}
  \caption{Degree Distribution of Tokens on the Uniswap V2 Market Network}
  \label{fig:Degree_Distribution}
\end{figure}

\begin{figure}[htbp]
  \centering
  \includegraphics[width=0.7\linewidth]{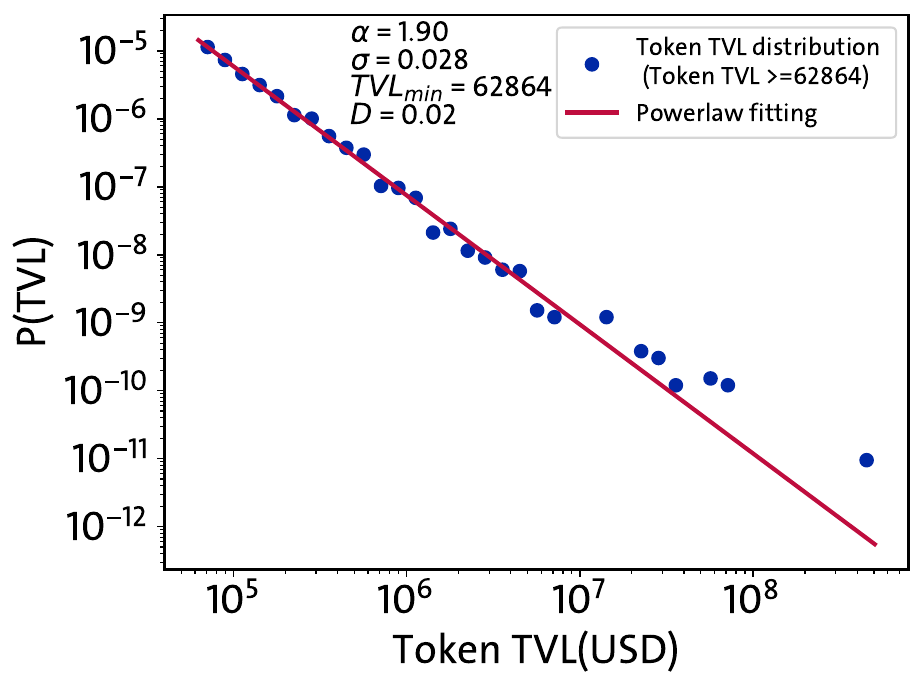}
  \caption{TVL Distribution of Tokens on the Uniswap V2 Market Network}
  \label{fig:token_tvl_distribution}
\end{figure}

\begin{figure}[htbp]
  \centering
  \includegraphics[width=0.7\linewidth]{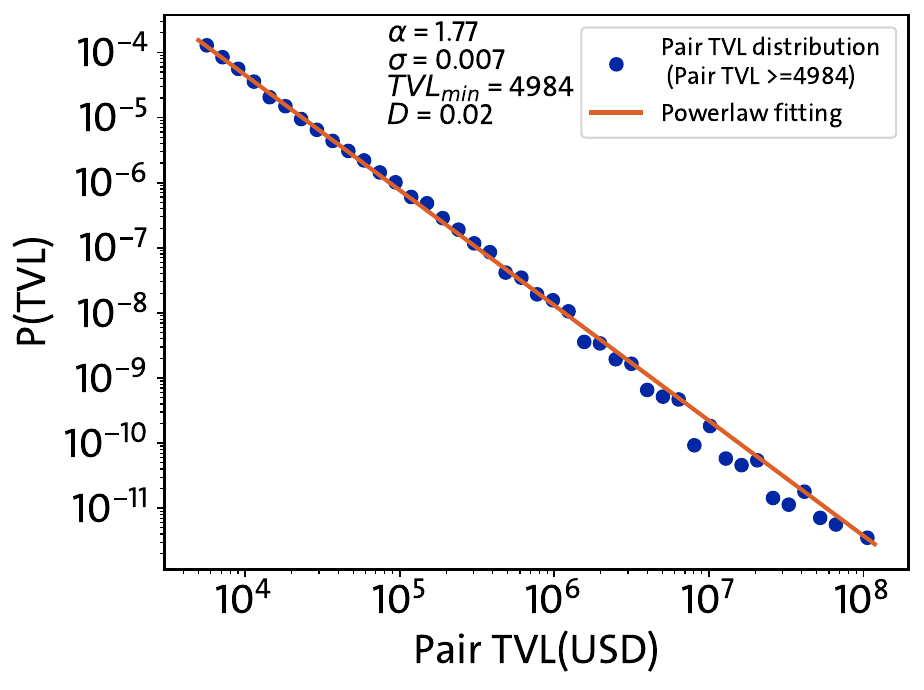}
  \caption{TVL Distribution of liquidity pools on the Uniswap V2 Market Network}
  \label{fig:edge_tvl_distribution}
\end{figure}

Similar to Figure \ref{fig:Degree_Distribution}, Figures \ref{fig:token_tvl_distribution} and \ref{fig:edge_tvl_distribution} display the power-law distribution of both liquidity pools and tokens' TVL, indicating that the distribution of TVL value tends to be concentrated on a few tokens and liquidity pools. Moreover, we have computed the Gini index for both tokens and liquidity pools' TVL, which reveals a high level of centralization. Specifically, the Gini index for tokens' TVL is 0.996, while the Gini index for liquidity pools' TVL is 0.99.

\subsection{Core-periphery structure}
A k-core refers to a subgraph that contains nodes with a degree of at least k. The core-periphery\cite{rombach2014core} structure refers to a network structure where a densely connected core of nodes interacts with less connected peripheral nodes. This hierarchical organization distinguishes central, highly interconnected elements (core) from less connected elements (periphery). Financial networks often exhibit a core-periphery structure\cite{campajola2022evolution, barucca2016disentangling}.

\begin{figure}[htbp]
  \centering
  \includegraphics[width=0.95\linewidth]{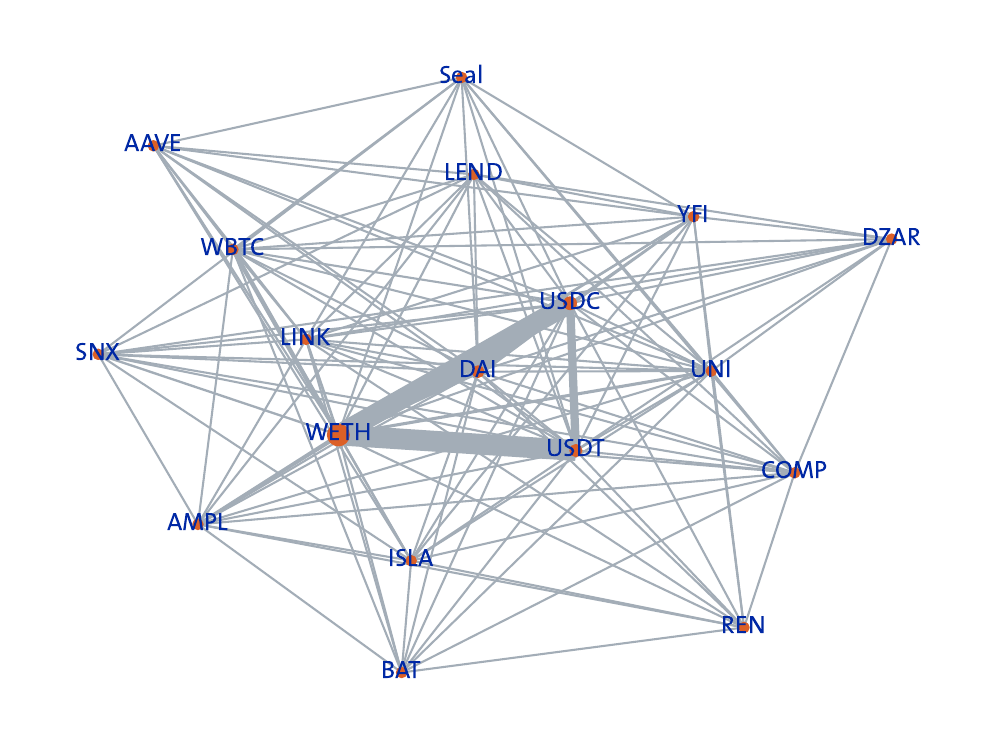}
  \caption{Visualization of the 11-core group. The size of nodes and the width of edges are proportional to the TVL of tokens and liquidity pools.}
  \label{fig:network_231031_k_core}
\end{figure}

\begin{figure}[htbp]
  \centering
  \includegraphics[width=0.8\linewidth]{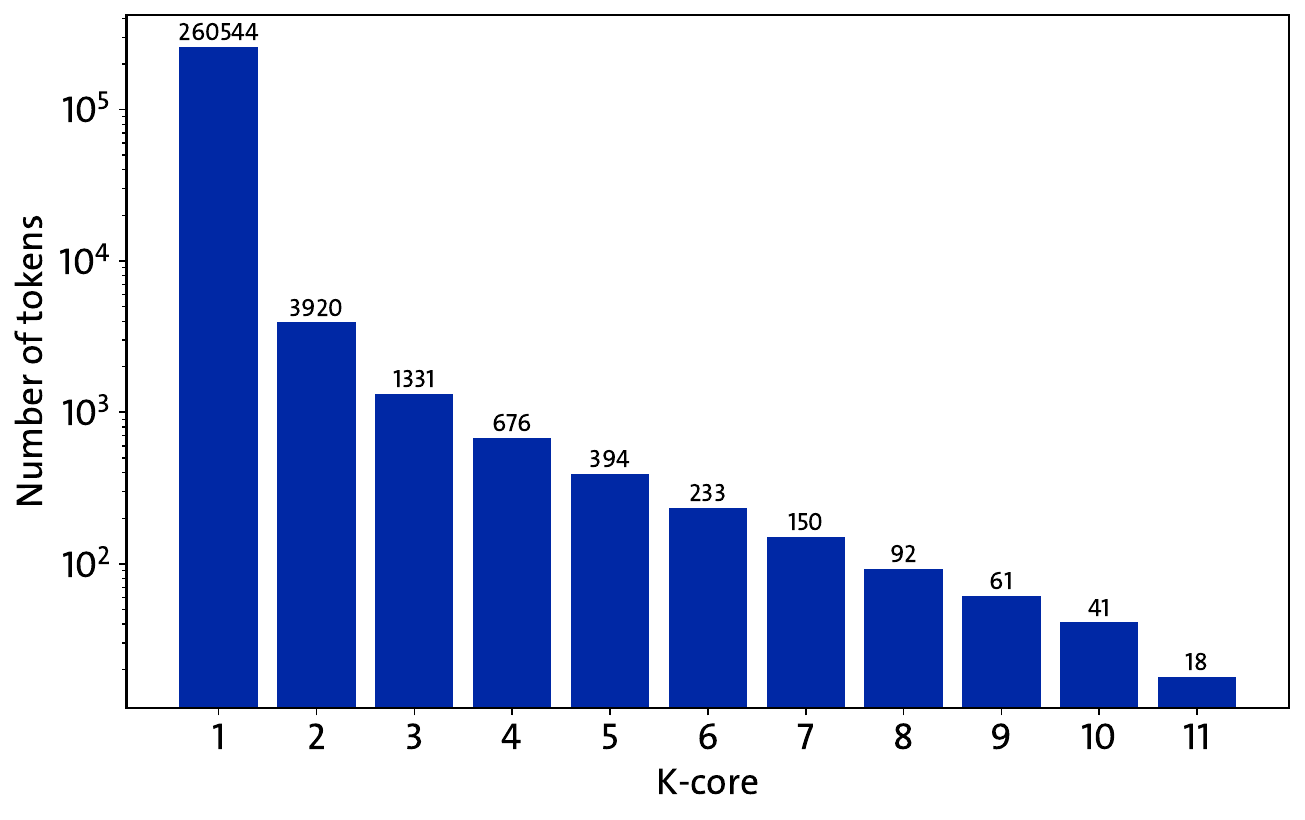}
  \caption{Number of tokens in different k-core groups}
  \label{fig:network_231031_k_core_number}
\end{figure}

In our analysis, the highest value of k observed is 11. Within this 11-core community, we identify 18 tokens and 125 liquidity pools. The visual representation of this community is depicted in Figure \ref{fig:network_231031_k_core} where the size of nodes and the width of edges are proportional to the TVL of tokens and liquidity pools. This 11-core group represents the most tightly interconnected community on Uniswap V2. Figure \ref{fig:network_231031_k_core_number} elaborates the number of nodes in different k-cores, showing that the size of the 2-core group is much smaller than the 1-core group, indicating that most tokens on Uniswap have only one liquidity pool. 

To determine whether this network has a significant core-periphery structure, we calculate and compare the k-core between the Uniswap network and a random network with the same number of nodes and edges. From table \ref{tab:network_comparison}, we observe that the random network has a 2-core group which consists of 128,500 nodes with an average degree of 2.72. In comparison, the average degree of the entire random network is 2.05. The number of nodes in the 11-core Uniswap group is much smaller than the 2-core group in the random network and the entire Uniswap network, while the average degree in the Uniswap 11-core group is much higher than the 2-core group in the random network and the entire Uniswap network, highlighting the presence of a core-periphery structure within the network.

\begin{table}[htbp]
    \renewcommand{\arraystretch}{1.3}
    \caption{Comparison between Uniswap Network and Random Network}
    \label{tab:network_comparison}
    \centering
    \begin{tabular}{lcc}
        \hline
        \textbf{Metrics} & \textbf{Uniswap Network} & \textbf{Random Network} \\
        \hline
        Number of Nodes & 260544 & 260544 \\
        Avg. Degree (Largest Component) & 2.05 & 2.05 \\
        K number in the k-core & 11 & 2 \\
        Number of Nodes (k-Core Group) & 18 & 128500 \\
        Avg. Degree (k-Core Group) & 6.94 & 2.72 \\
        \hline
    \end{tabular}
\end{table}

\subsection{Token and Liquidity Pool Importance}
In this study, we employ the concept of betweenness as a metric for assessing the significance of tokens and liquidity pools within the network. We choose to use betweenness because it measures the number of times a node or an edge acts as a bridge along the shortest path between two other nodes. In our case, it measures the number of times a token or liquidity pool acts as a bridge between two other tokens or liquidity pools in the network. A node with a high betweenness value has a significant influence on the flow of tokens in the network. 
\begin{table}[htb]
\caption{Token Betweenness}
\label{tab:node_betweenness_info}
\centering
\begin{tabular}{llc}
\hline
\textbf{Token} & \textbf{Address} & \textbf{Weight} \\
\hline
WETH  & 0xc02aaa39b223fe8d0a0e5c4f27ead9083c756cc2 & 0.999495 \\
USDC  & 0xa0b86991c6218b36c1d19d4a2e9eb0ce3606eb48 & 0.015468 \\
USDT  & 0xdac17f958d2ee523a2206206994597c13d831ec7 & 0.009354 \\
DAI   & 0x6b175474e89094c44da98b954eedeac495271d0f & 0.002833 \\
WBTC  & 0x2260fac5e5542a773aa44fbcfedf7c193bc2c599 & 0.000423 \\
PEPE  & 0x6982508145454ce325ddbe47a25d4ec3d2311933 & 0.000393 \\
SHIB  & 0x95ad61b0a150d79219dcf64e1e6cc01f0b64c4ce & 0.000235 \\
DEXTF & 0x5f64ab1544d28732f0a24f4713c2c8ec0da089f0 & 0.000234 \\
UNI   & 0x1f9840a85d5af5bf1d1762f925bdaddc4201f984 & 0.000210 \\
TSUKA & 0xc5fb36dd2fb59d3b98deff88425a3f425ee469ed & 0.000173 \\
\hline
\end{tabular}
\end{table}

\begin{figure}[htbp]
  \centering
  \includegraphics[width=0.9\linewidth]{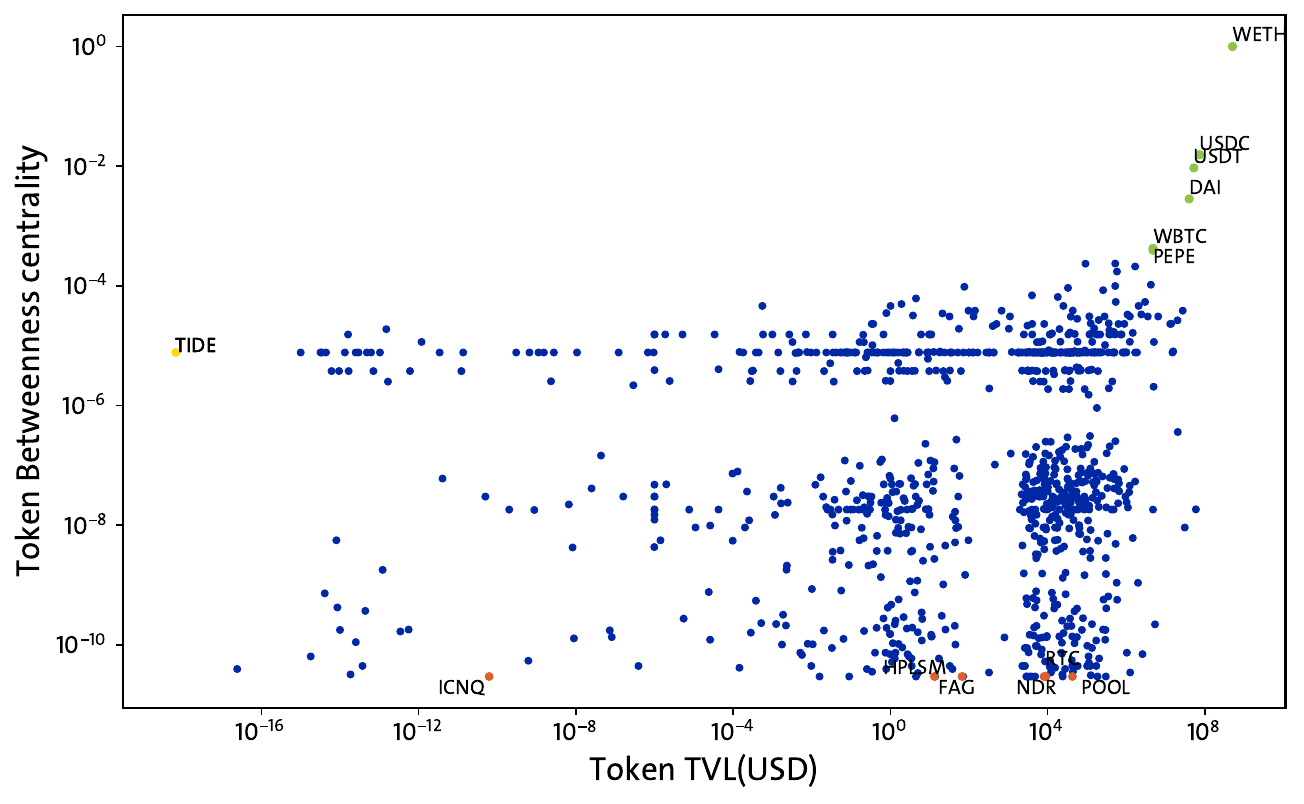}
  \caption{Token TVL and token betweenness}
  \label{fig:token_between}
\end{figure}

By using betweenness as a node importance measurement, we compute the betweenness centrality for each token. Table \ref{tab:node_betweenness_info} presents a comprehensive list of the top 10 tokens exhibiting the highest betweenness centrality. These tokens are highly influential in the network as they act as bridges between other tokens and liquidity pools. Figure \ref{fig:token_between} reveals an interesting observation where the overall betweenness of tokens seems to demonstrate no significant correlation with the Total Value Locked (TVL) of nodes. Nevertheless, it is noteworthy that among the top 6 tokens based on TVL, they exhibit the highest betweenness as well. Furthermore, we note the presence of certain tokens, HPLSM and PAG, which possess a relatively high TVL despite having a relatively small betweenness value.

\begin{figure}[htbp]
  \centering
  \includegraphics[width=0.9\linewidth]{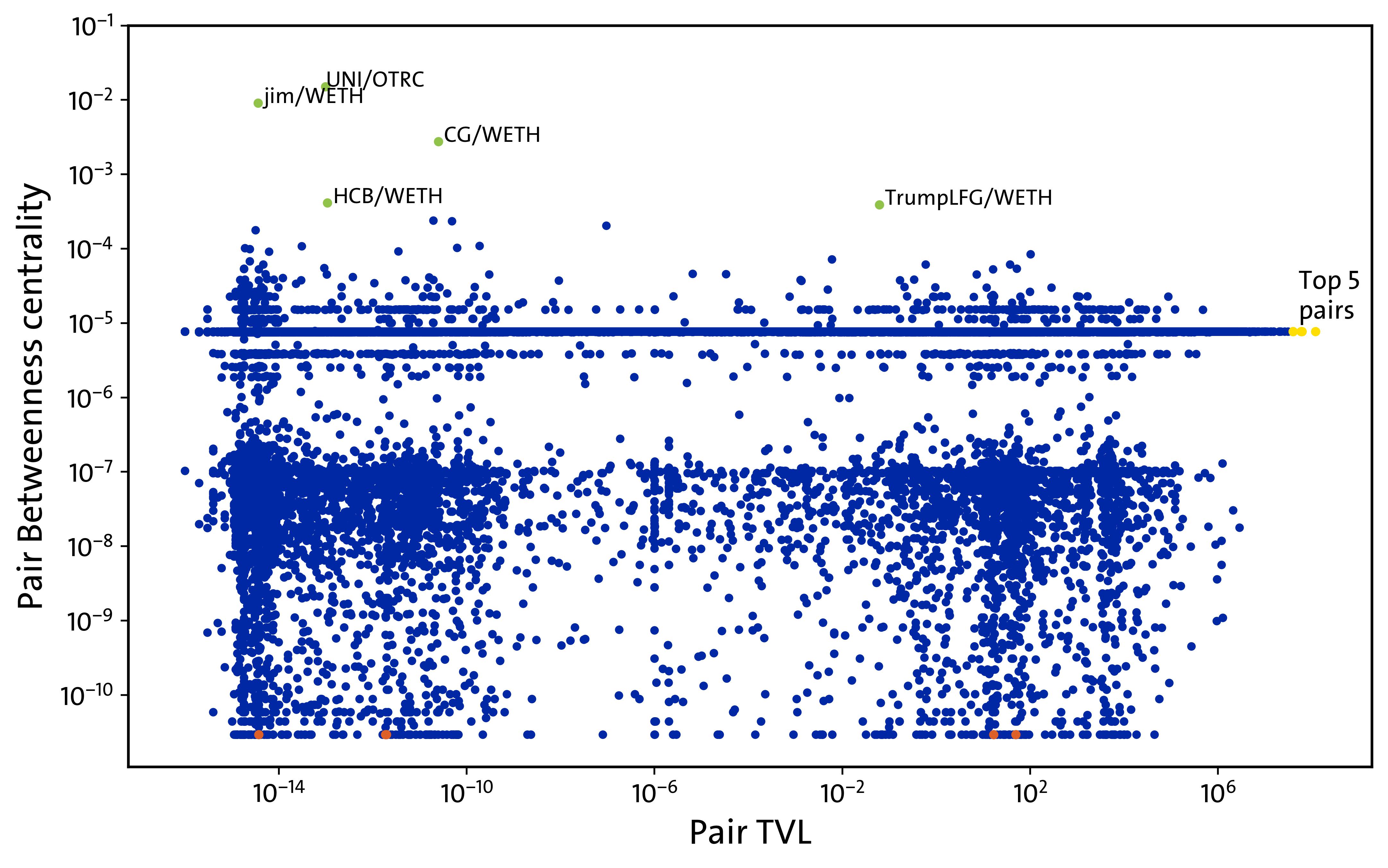}
  \caption{Pair TVL and pair betweenness centrality}
  \label{fig:edge_between}
\end{figure}
Table \ref{tab:edge_betweenness_info} presents a comprehensive list of the top 10 liquidity pools exhibiting the highest betweenness centrality. Interestingly, it is observed in Figure \ref{fig:edge_between} that all of these pairs display a significantly low Total Value Locked (TVL). Nonetheless, their abundance in shortest paths within the network structure suggests their crucial role. The representation of this information can be visualized in Figure \ref{fig:edge_between}, where the liquidity pools exhibiting the greatest betweenness are depicted in green, whereas the five pools with the lowest betweenness centrality are displayed in red. Remarkably, the top five liquidity pools in yellow color with the highest Total Value Locked (TVL) do not necessarily possess the highest betweenness centrality values.

\begin{table}[tbh]
\caption{Liquidity Pool (Edge) Betweenness}
\label{tab:edge_betweenness_info}
\centering
\begin{tabular}{llc}
\hline
\textbf{Pair Name} & \textbf{Pair Address} & \textbf{Betweenness Centrality} \\
\hline
UNI/OTRC           & 0x3e18ef2208e159b9e87130eb7278507eedb26b18  & $0.015149$ \\
jim/WETH           & 0x034beaeb9adaa6366abf49a6a1aad2d920edeb15  & $0.009108$ \\
CG/WETH            & 0x3ea17f74db2cc487b6641275cddbda92afaa9b37  & $0.002765$ \\
HCB/WETH           & 0x3f0f78d50721e0f499226d0cd8dc89f80acd0dc1  & $0.000415$ \\
TrumpLFG/WETH      & 0x41bfc675d76b1d7dbb6ce245f289a62c85fdb571  & $0.000392$ \\
SHIBAKIBA/WETH     & 0x4225af5d2a6d67fbebe5a5283a58fd474abc6ffa  & $0.000240$ \\
HAY/WETH           & 0x4145d769c6cbfbfec8474745580a16fe406b5535  & $0.000235$ \\
MSM/WETH           & 0x3fd01436040d1f61f2d9fc57868dc406e35da08a  & $0.000205$ \\
WETH/CTQ           & 0x26b2fde5c369e16f02a02d9b27d7aba5ccbb0515  & $0.000178$ \\
WETH/GIZMO         & 0x458e44fd9d4ef01077cf3dcf42fa49f2f76573f3  & $0.000109$ \\
\hline
\end{tabular}
\end{table}

\section{Dynamic feature of the Uniswap Network}

We calculate the dynamic network features by constructing the daily network, each daily network contains all the liquidity pools that have been created up to that specific day. Given the perceived fragility of the current network, our objective is to ascertain the evolution of the fragility of the Uniswap network. To investigate the dynamic fragility of the system, we assess various indicators, including network size, component number, scale-free property and core-periphery property. 
\subsection{Network size evolution}
\begin{figure}[htbp]
  \centering
  \includegraphics[width=0.8\linewidth]{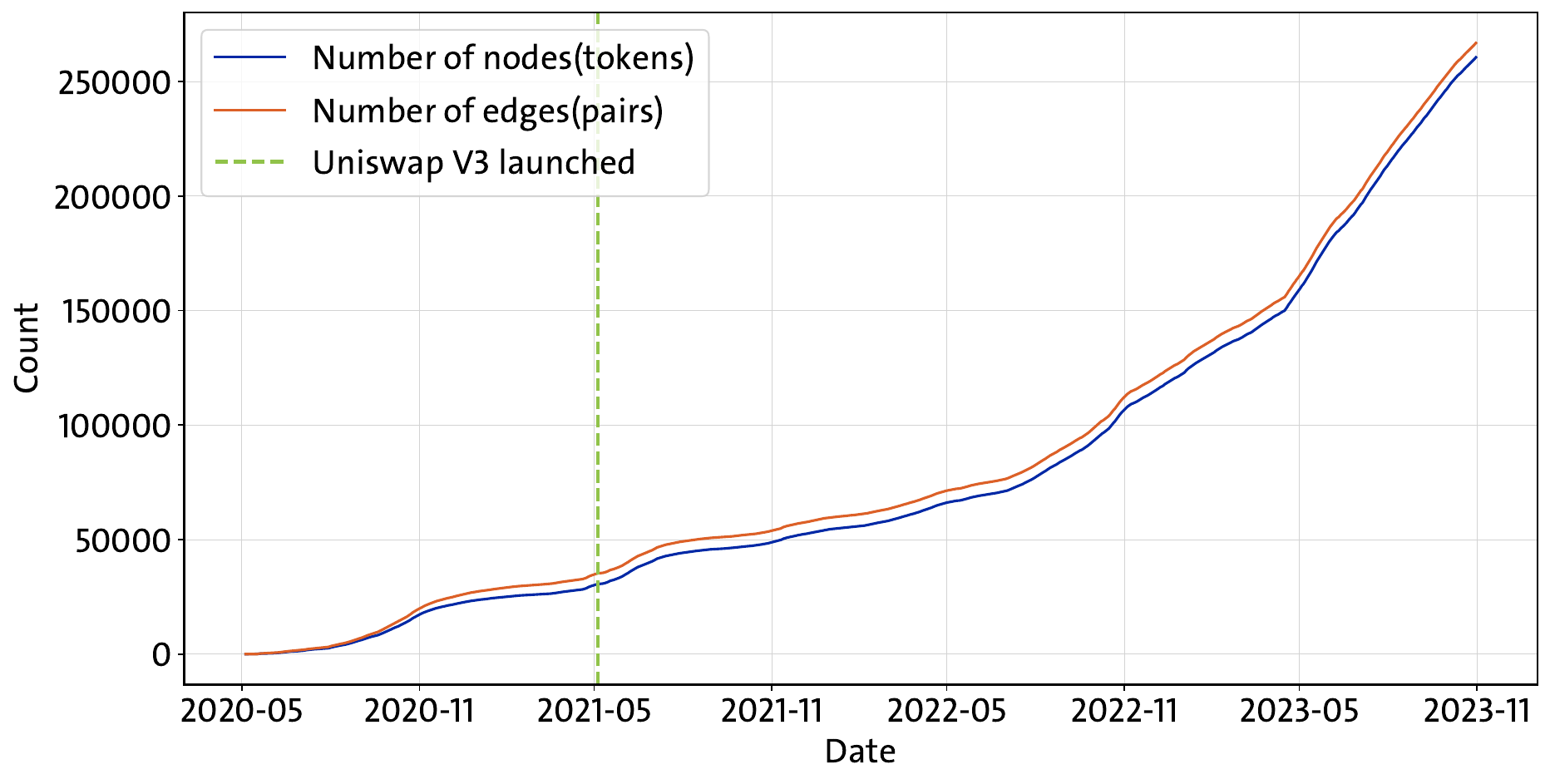}
  \caption{The size of Uniswap network over time}
  \label{fig:number_nodes_edges}
\end{figure}

Depicted in Figure \ref{fig:number_nodes_edges}, the network has shown a clear growth trend from the initial 2 nodes and 1 edge on May 5, 2020 to the latest count of 260,544 nodes and 266,826 edges on October 31, 2023. Notably, there has been a significant liquidity pool created in the network although Uniswap V3 was launched on May 5, 2021. Moreover, the number of liquidity pools consistently approximated the count of tokens, indicating this is a sparse network and the average degree is around one. 

\subsection{Network components}
\begin{figure}[htbp]
  \centering
  \includegraphics[width=0.8\linewidth]{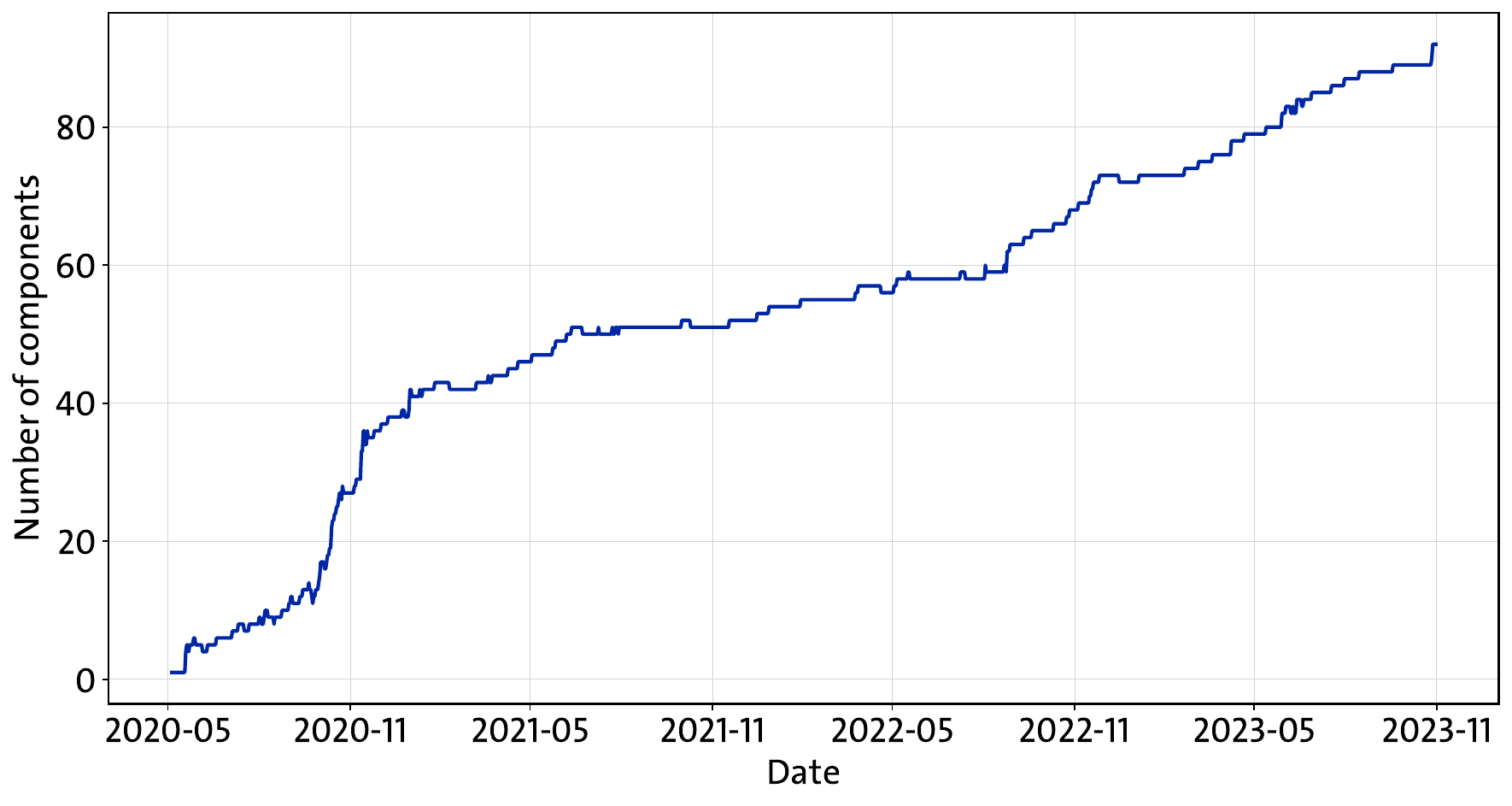}
  \caption{Number of components}
  \label{fig:number_components}
\end{figure}

The number of network components can be used to measure the connectivity in the Uniswap market. A higher number of components indicates a lower level of network connectivity. Ideally, the Uniswap platform should operate as a connected network where any two tokens can be traded seamlessly. However, the actual Uniswap network consists of multiple components, which restrict the ability to freely trade every token with any other token. Figure \ref{fig:number_components} shows the evolution of the components within the Uniswap network over time. The findings indicate a consistent rise in the number of components, suggesting a growing fragmentation within the network. As a result, there is an increasing prevalence of token groups that are isolated, preventing them from trading with tokens belonging to other groups. This fragmentation adversely affects the overall efficiency of the Uniswap market. However, there are occasional slight decreases in the number of components, which can be interpreted as a positive outcome. It indicates that network connectivity is enhanced through the establishment of some new liquidity pools. 

\subsection{Scale-free property}
\begin{figure}[htbp]
  \centering
  \includegraphics[width=0.8\linewidth]{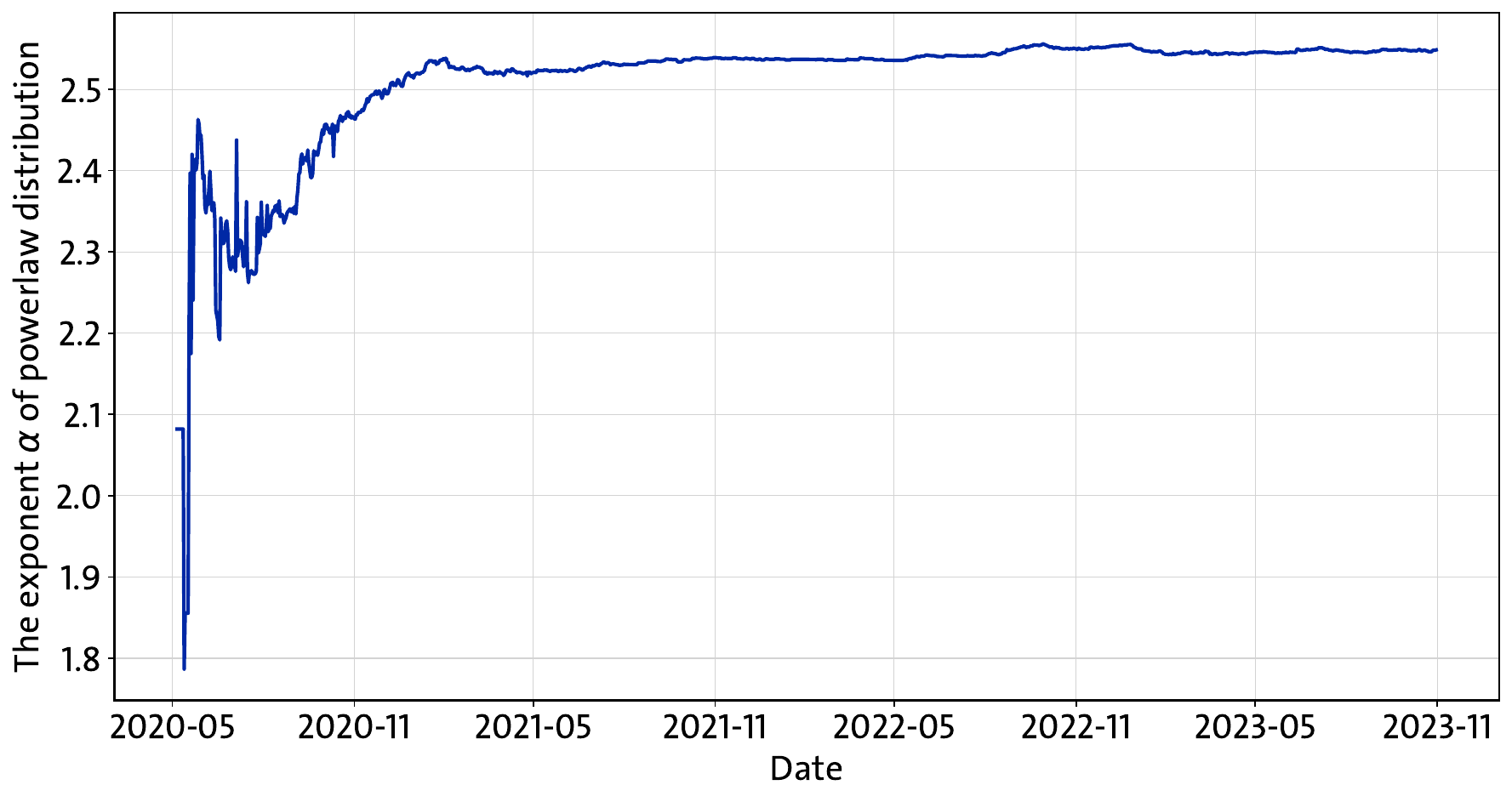}
  \caption{The value of ${\alpha}$ in the power-law distribution of daily networks}
  \label{fig:powerlaw_alpha}
\end{figure}

According to Figure \ref{fig:powerlaw_alpha}, the power-law exponent ${\alpha}$ exhibits greater variability during the initial stages of the Uniswap network and steadily increases until February 2021. Subsequently, there is a minor upward trend observed until October 31, 2023. As explained in Section \ref{Method}, a higher value of ${\alpha}$ signifies a higher concentration of degrees within a relatively smaller proportion of tokens, indicating a growing level of network fragility.
\subsection{Core-periphery structure}
\begin{figure}[htbp]
  \centering
  \includegraphics[width=0.8\linewidth]{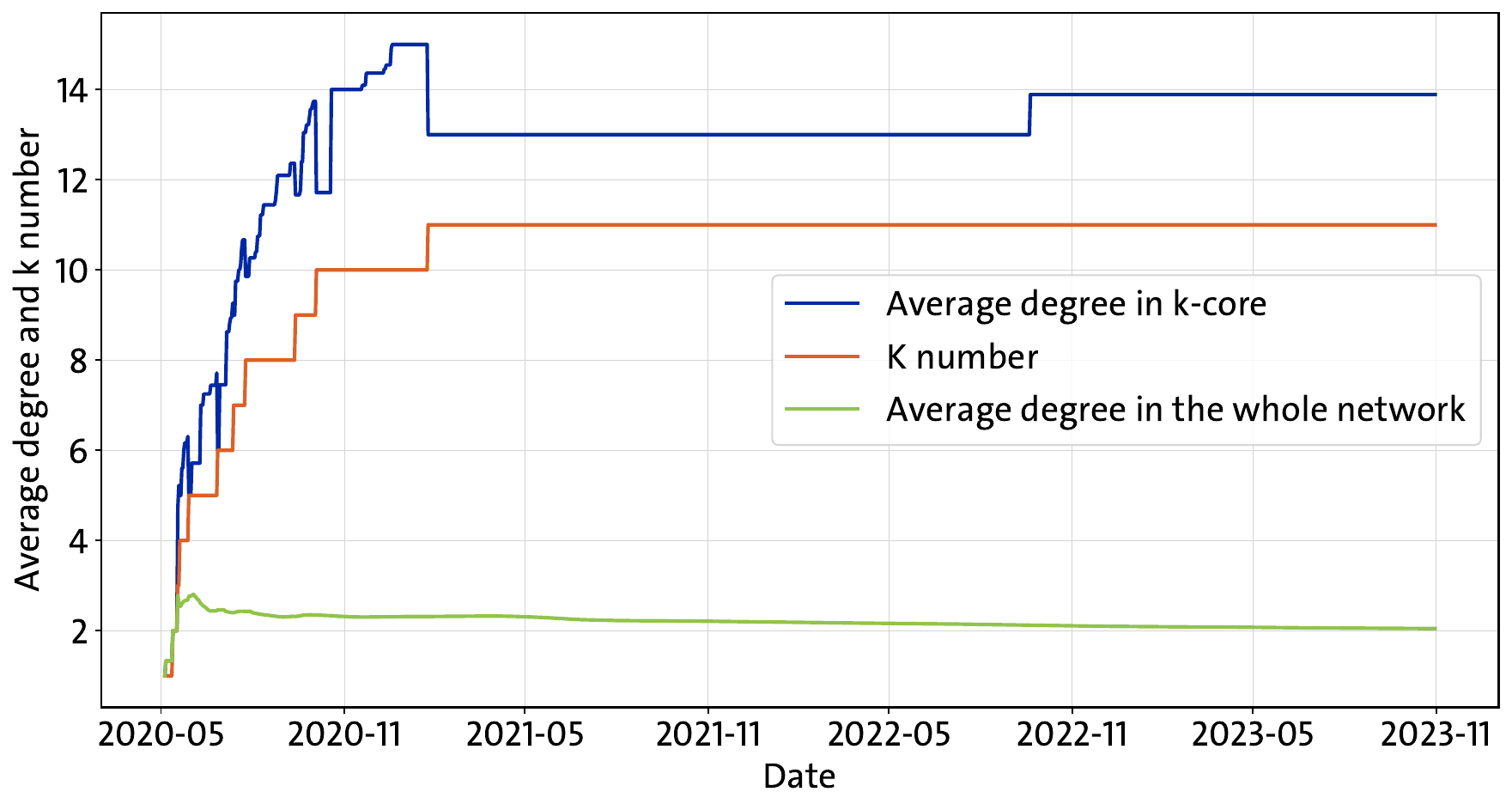}
  \caption{Average degree in the k-core and the entire network}
  \label{fig:average_degree_k}
\end{figure}

\begin{figure}[htbp]
  \centering
  \includegraphics[width=0.8\linewidth]{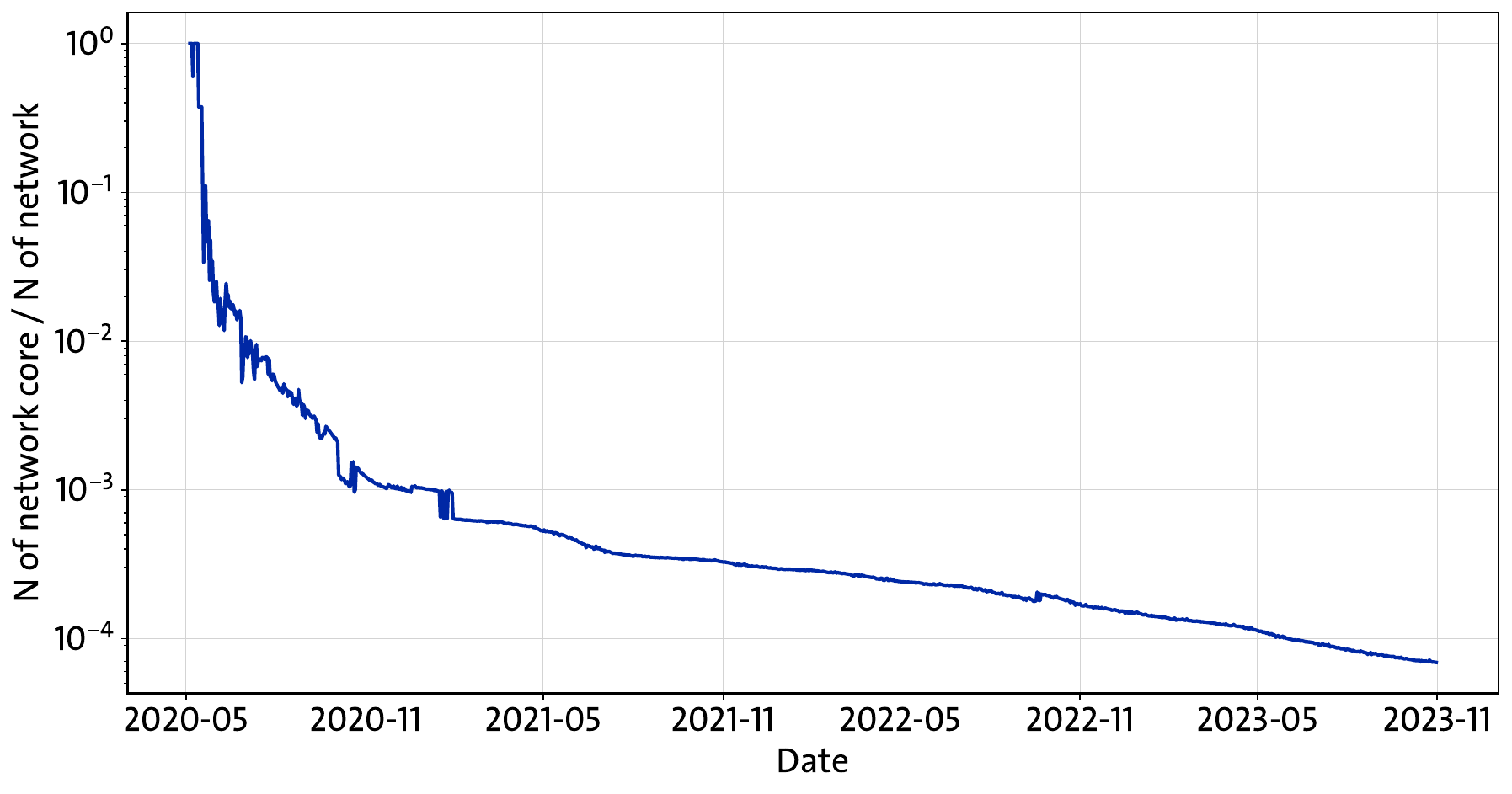}
  \caption{The ratio: number of nodes in k-core / number of tokens in the entire network}
  \label{fig:k_core_nodes_ratio}
\end{figure}

Figures \ref{fig:average_degree_k} depict the average degrees of the overall network and its k-core, respectively. It is evident that the average degree within the network core exceeds that of the entire network. Similarly, the proportion of nodes within the network core diminishes in Figure \ref{fig:k_core_nodes_ratio}, suggesting the existence of a few tightly connected groups within this network. This delineates a core-periphery structure of the Uniswap network, which is steadily gaining prominence.

In summary, these dynamic network features indicate that the Uniswap network becomes increasingly fragile over time. This is observed through its low density, higher number of components, and the growing value of ${\alpha}$ in the power-law distribution, along with an increasingly significant core-periphery structure.  

\section{Robustness Analysis}

\subsection{Token collapse effect}
From the static and dynamic feature analysis we find that the Uniswap network is a scale-free network; from the study\cite{doyle2005robust, hasheminezhad2020scale}, scale-free networks show a 'robust yet fragile' feature due to their heterogeneous degree distributions. Such networks are more likely to remain connected than random 
networks following the removal of randomly selected nodes, but they are also more
susceptible to targeted attacks.

In May 2022, TerraUSD (UST), the fourth-largest stablecoin, experienced a detrimental collapse due to the loss of its peg to the U.S. dollar. Recent research \cite{badev2023interconnected,lee2023dissecting} demonstrates that this collapse significantly affected the interconnectedness of the cryptocurrency market. If another token were to undergo a collapse similar to Terra's, it would also impact the Uniswap market. We assume that on Uniswap, once a token collapses, all liquidity pools paired with that token will lose their value and will no longer be traded. This assumption is reasonable because the collapsed token will lose its value on all cryptocurrency markets, leading individuals to exchange their collapsed tokens for another token from the liquidity pool. While the price of the collapsed token may be very low in the pool, users will not swap it for another token. As a result, the liquidity pools paired with the collapsed token will not be traded on Uniswap. Therefore, we can consider all connections linked to the collapsed token as deleted or removed, which will impact the connectivity and total value locked (TVL) of the Uniswap market.

\subsection{Removing methods}
We utilize four methods for token deletion: deletion by the TVL, deletion by betweenness centrality, deletion by degree centrality, and random deletion as a comparative measure. These three indicators are ranked in descending order. The network's robustness is measured using three metrics: the number of remaining liquidity pools (edges), the number of persistent components, and the proportion of initial TVL lost calculated after each node removal. Subsequently, we delete the first 1,000 tokens complying with 4 kinds of orders, It is worth mentioning that the WETH, USDT, USDC, and DAI tokens are excluded due to their comprehensive mechanisms and strong endorsements.

\subsection{Result analysis}
\begin{figure}[htbp]
  \centering
  \includegraphics[width=0.95\linewidth]{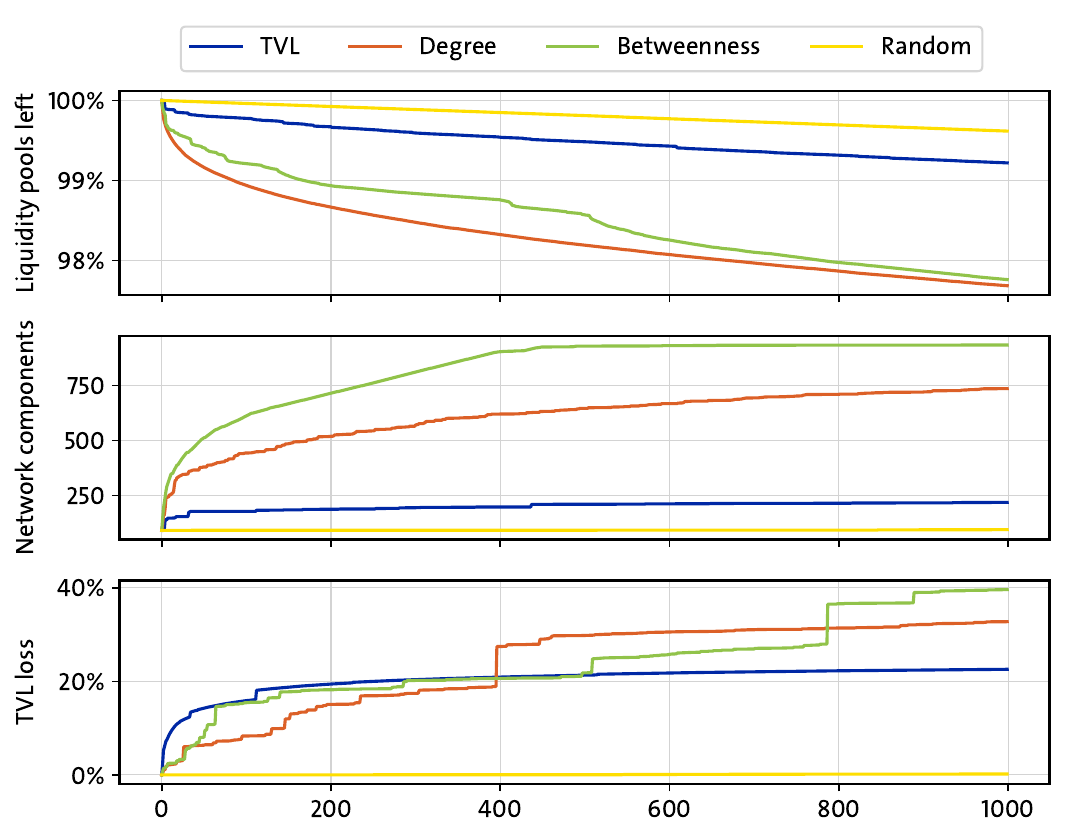}
  \caption{Robustness of the Uniswap network. This analysis employs four methods for node removal: sorting by TVL, sorting by degree, sorting by betweenness, and random deletion. Three metrics are used to measure the impact after removal: the remaining number of edges, the number of components formed, and the amount of TVL lost.}
  \label{fig:robustness_analysis}
\end{figure}

The impact of node removal is illustrated in Figure \ref{fig:robustness_analysis}. We investigate the results of different node removal strategies on overall network performance. Comparing the effect of random removal, we observe that this network demonstrates robustness, as the random selection of nodes has minimal influence on the number of liquidity pools and network components, as well as the TVL loss. Nevertheless, this scenario alters when nodes are removed using the other three alternative methods. In the subsequent sections, we investigate the effects of the other three node removal methods in detail.
\subsubsection{Liquidity pools remained} 
The number of remaining liquidity pools is minimized when tokens are sorted and removed based on the degree centrality. This is because nodes with higher degrees tend to have more connections with other nodes, and therefore their removal leads to a disruption in the network's connectivity. Following the removal based on betweenness centrality, the number of remaining liquidity pools is second only to the degree centrality approach, and the results are very close when the number of nodes reaches 800. This indicates that nodes with higher betweenness centrality also have many connections with other nodes.

\subsubsection{Components generated} 
The number of components formed is maximized when nodes are sorted and removed based on betweenness centrality and significantly surpasses the number obtained using other removal methods. This is because nodes with higher betweenness centrality are important in transmitting information within the network, and removing them based on betweenness centrality results in a more dispersed network.

\subsubsection{Liquidity lost} 
The amount of Total Value Locked (TVL) lost varies with different node removal methods as the number of nodes removed changes. Prior to removing approximately 400 tokens, removing nodes based on TVL sorting causes the largest loss of liquidity, followed by removal based on betweenness centrality sorting. After this point, removing nodes based on degree sorting results in the highest TVL loss, which continues until approximately 800 nodes. Finally, the removal based on betweenness leads to the most significant loss of TVL compared to the removal based on TVL and degree sorting.

In summary, the analysis results demonstrate that the performances of different removal methods differ on different metrics. Removing nodes based on degree results in the fewest remaining liquidity pools, while removing them based on betweenness leads to the highest number of isolated components. Removing nodes based on TVL sorting results in the fastest decline in Uniswap's liquidity (in terms of removing 400 nodes). However, overall, based on the performance of the four removal methods across three different metrics, removing nodes based on betweenness has a greater impact on the network. 

\section{Conclusion}
In this paper, we conduct a comprehensive analysis of the Uniswap V2 market network from the perspective of a complex network. We construct the network by using the liquidity pool data and token TVL data and analyze its static and dynamic features. Additionally, the study delves into the network's robustness by exploring the effects of various node removal methods. The findings suggest that the Uniswap V2 market network adopts a scale-free property with a core-periphery structure. Moreover, the network demonstrates increasing fragility over time. While it exhibits robustness against random node removal, it is vulnerable to the removal of nodes with high betweenness centrality. Overall, this study reveals that although Uniswap is a decentralized exchange, it exhibits significant centralization characteristics in terms of token network connectivity and the TVL distribution of nodes(tokens) and edges(liquidity pools). There are still several areas of further research that can be pursued. For instance, investigating alternative metrics to explore the characteristics of networks and measure the importance of tokens and liquidity pools, utilizing epidemiological models\cite{keeling2005networks} to study how risk propagates within networks if certain tokens crash. Additionally, one could conduct robustness analysis by removing liquidity pools (edges) subjected to attacks such as the Rug-pull attack\cite{mazorra2022not} and verifying the connectivity of the Uniswap network.



\vspace{12pt}


\end{document}